\documentclass[structabstract]{aa}
\usepackage{graphicx}
\usepackage{txfonts}
\usepackage{natbib}

\bibpunct[]{(}{)}{;}{a}{}{,}
\begin{document}
   \title{Accurate laboratory rest frequencies of vibrationally
          excited CO up to $\varv = 3$ and up to 2~THz}

   \author{R. Gendriesch\inst{1}
           \and
           F. Lewen\inst{1}
           \and
           G. Klapper\inst{1}
           \and
           K.~M. Menten\inst{2}
           \and
           G. Winnewisser\inst{1}
           \and
           J.~A. Coxon\inst{3}
           \and
           H.~S.~P. M{\"u}ller\inst{1}
          }

   \institute{I.~Physikalisches Institut, Universit{\"a}t zu K{\"o}ln,
              Z{\"u}lpicher Str. 77, 50937 K{\"o}ln, Germany\\
              \email{lewen,winnewisser,hspm@ph1.uni-koeln.de}
         \and
              Max-Planck Institut f\"ur Radioastronomie, Auf dem H\"ugel 69,
              53121 Bonn, Germany\\
              \email{kmenten@mpifr-bonn.mpg.de}
         \and
              Department of Chemistry, Dalhousie University, 
              Halifax, Nova Scotia, B3H 4J3, Canada
              \email{John.Coxon@Dal.Ca}
             }

   \date{Received 18 December 2008 / Accepted 06 February 2009}

  \abstract
{}
{Astronomical observations of (sub)millimeter wavelength pure rotational 
 emission lines of the second most abundant molecule in the Universe, CO, 
 hold the promise of probing regions of high temperature and density in 
 the innermost parts of circumstellar envelopes.}
{The rotational spectrum of vibrationally excited CO up to $\varv = 3$ 
has been measured in the laboratory between 220 and 1940~GHz with 
relative accuracies up to $5.2 \times 10^{-9}$, corresponding to 
$\sim 5$~kHz near 1~THz.}
{The rotational constant $B$ and the quartic distortion parameter $D$ have
been determined with high accuracy and even the sextic distortion term $H$
was determined quite well for $\varv = 1$ while reasonable estimates of $H$ 
were obtained for $\varv = 2$ and 3.}
{The present data set allows for the prediction of accurate rest frequencies
of vibrationally excited CO well beyond 2~THz.}


\titlerunning{rest frequencies of excited CO}

\maketitle
\hyphenation{For-schungs-ge-mein-schaft}

%

\section{Introduction}
\label{intro}

Carbon monoxide, CO, is the second most abundant molecule in the
universe next to hydrogen, H$_2$. Observations of CO are commonly
employed to obtain information on the H$_2$ abundance since H$_2$
is usually difficult to detect directly because it does not have
any permanent or transition dipole moment, and its quadrupole
transitions are very weak and high in energies.
Usually transitions of the main isotopologue $^{12}$C$^{16}$O
(in the following, unlabeled atoms designate $^{12}$C or $^{16}$O)
are recorded for this purpose, but frequently also transitions of
$^{13}$CO or even C$^{18}$O are studied because transitions of
these isotopic species are less affected by opacity.
Even less abundant isotopologues have been detected in space,
including recently the least abundant $^{13}$C$^{17}$O 
\citep{Det_13C17O}.

Very accurate ground state rest frequencies have been obtained
in the laboratory for CO \citep{CO_v0_rot_1997} as well as for
its isotopologues
\citep{1316_TuFIR,1316_GK,1318_GK,1218_GK,1217_1317_GK,1217_GC,1218_GC,1316_GC,1317_1318_CP}.
The lower-$J$ transitions were usually measured in sub-Doppler
resolution such that the relative accuracies of the measurements
reach $\sim 10^{-9}$. 

\begin{figure}
\includegraphics[angle=-90,width=9cm]{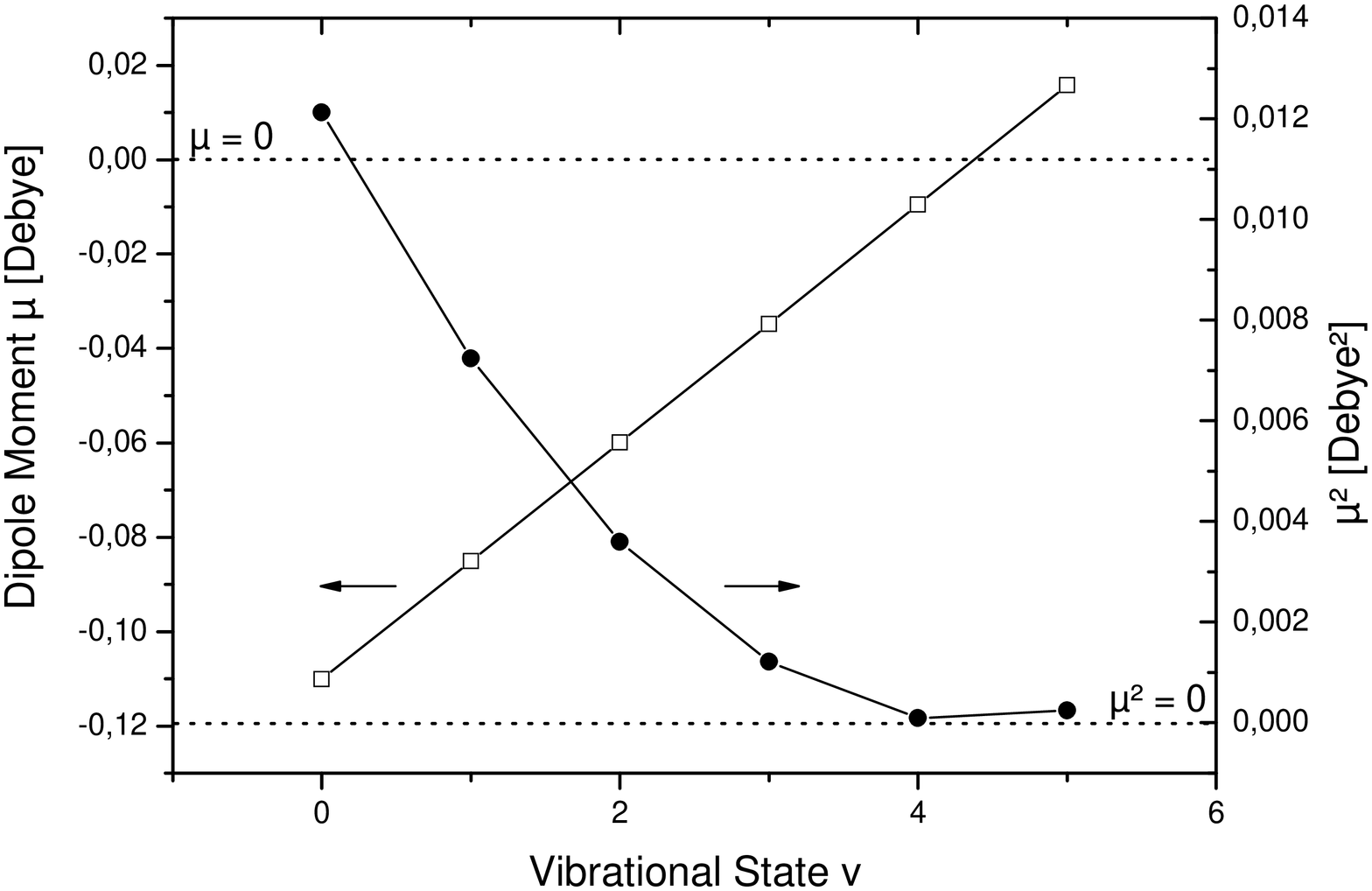}
  \caption{The open squares with the $y$-scale to the left show the 
decreasing dipole moment from $\varv = 0$ to 4. Between $\varv = 4$ 
and 5 the dipole moment switches polarity from the negative polarity 
being at the C atom for $\varv \leq 4$ to the negative polarity 
being at the O atom for $\varv \geq 5$. The filled circles show the much 
larger effect in the squares of the dipole moments which are proportional 
to the intensities of the lines. The $y$-scale is on the right hand side.}
\label{dip}
\end{figure}

Interestingly, the only experimental article to report measurements
of transitions of vibrationally excited CO seems to be a paper by
\citet{CO_vib_J_2-1} on CO $J = 2 - 1$ transitions for several
vibrational states between $\varv = 5$ and $\varv = 40$.
One reason for this lack of measurements is probably the high
vibrational energy of 2143~cm$^{-1}$ or 3086~K for $\varv = 1$
\citep{CO_pot_2004} which makes excited states of CO difficult
to populate.
Another reason is likely the small dipole moment of 0.10980~(3)~D
\citep{CO_Stark} which even decreases for the first few vibrational
states by about 0.025~D for each additional vibrational excitation
until the dipole moment switches polarity between $\varv  = 4$ and 5
\citep{CO_dipole_surface}, see also Fig~\ref{dip}.
\citet{CO_pot_2004} reviewed rather extensively not only rotational 
data but also a plethora of infrared studies. Their work permits 
very good predictions of the rotational transition of CO in excited 
vibrational states.


Apart from molecular clouds, CO is also a premier constituent of the 
circumstellar envelopes of mass-losing stars on the asymptotic giant 
branch (AGB) of both carbon-rich ([C/O] abundance ratio $> 1$) and 
oxygen-rich ([C/O] $< 1$) objects. Moreover, unlike other species, 
it does not condense into dust grains and prevails throughout the 
envelope until it is destroyed by ultraviolet radiation in its tenuous 
outer portions \citep[see, e.g., ][]{Lafont1982, Cherchneff1992}. 
Thus, emission from its strong rotational lines dominate the cooling 
and, thus, the thermal balance of an envelope 
\citep[see, e.g. ][]{CrosasMenten1997}. 

CO is produced abundantly under the thermodynamical equilibrium conditions 
in the atmospheres of such AGB stars, even at photospheric temperatures, 
$\sim 2000$ K, \citep{Tsuji1964} and is consequently observed easily via 
its near-infrared vibrational-rotational transitions. 
Toward the archetypical, high mass-loss C-rich object IRC+10216, 
\citet{SahaiWannier1985} observed vibration-rotational emission and 
absorption in the $4.6~\mu$m fundamental band, while \citet{Keady1988} 
additionally observed absorption from the first overtone band at $2.3~\mu$m.

Given the extremely high temperature \textit{and} density in the region 
from which the above transitions are arising, one might also expect 
to find pure rotational emission to be observable from it. 
However, pure rotational transitions of CO in excited vibrational states 
had not been reported in the astronomical literature until very recently 
(\citet{Patel2009a}, see also below), while rotational lines of molecules 
chemically related to CO, although with considerably lower vibrational 
energies, $E_{\rm vib}$, have been found toward IRC+10216. 
These include CS \citep{CS_vib} and SiS \citep{SiS_vib}, which have 
$E_{\rm vib} = 1272$~cm$^{-1}$  \citep{CDMS_2} and 744~cm$^{-1}$ 
\citep{SiS_2007}, respectively; see also \citep{Patel2009b}. 
Moreover, vibrationally excited SiO \textit{maser} emission has been detected 
from the envelopes of many hundreds of oxygen-rich long period variable stars 
and red supergiants, and both its $\varv = 1$ and 2 ($E_{\rm vib} = 
1230$ and 2447~cm$^{-1}$, respectively \citep{SiO_2003}) $J=1 - 0$ lines 
are observed even toward three high-mass protostellar objects embedded 
in molecular clouds \citep{Zapata2009}.

It is noteworthy that for all three of these molecules
transitions of even higher excited vibrational states and rare isotopologues 
have been observed, namely $\varv = 4$ for SiO \citep{SiO_v_4} and SiS
\citep{SiS_v_4,Patel2009b}, and $\varv = 3$ for CS \citep{CS_v_3} and
$^{29}$SiO \citep{29SiO_v_3}.

In the case of CO, \citep{ScovilleSolomon1978} reported the detection 
of its $\varv = 1$, $J = 1 - 0$ transition toward IRC+10216. 
However, that ``detection'' was soon rebutted by \citet{Cummins1980}. 
They identified the observed spectral feature as due to C$_4$H, whose 
frequency is very close indeed to that of the $\varv = 1$, $J = 1 - 0$ 
CO line. Moreover, C$_4$H exhibits generally strong lines in this source. 
It took roughly 30 years until a genuine detection of pure rotational 
CO emission lines ($\varv =1$, $J = 2 - 1$ and $3 -2$) with the 
Submillimeter Array was reported (of course in IRC+10216) very recently 
by \citet{Patel2009a}.

We have recorded rotational transitions of vibrationally excited CO
belonging to four isotopic species, CO, $^{13}$CO, C$^{18}$O, and
$^{13}$C$^{18}$O up to $\varv  = 3$ in excitation and up to 2~THz in
frequency. In order to facilitate the detection of radio lines
of vibrationally excited CO, we present here our results for the 
main CO isotoplogue. The full account of our work, which involves a 
combined fit with a variety of rotational and ro-vibrational data, 
will be presented elsewhere (Gendriesch et al., in preparation).

\section{Experimental details}
\label{exptl}

Transitions with frequencies below 1~THz have been recorded with the
Cologne Terahertz Spectrometer (CTS) which has been described in detail
by \citet{BWO-THz_spec}. It uses broadband tunable, phase-locked
backward-wave oscillators (BWOs) as powerful sources, and a magnetically
tuned, liquid helium cooled hot-electron InSb bolometer as detector.
The Cologne Sideband Spectrometer for Terahertz Application (COSSTA)
was employed for the measurements between 1.75 and 2.0~THz.
In this case, the terahertz radiation was generated by mixing of a
fixed-frequency far-infrared laser and a BWO as tunable sideband
generator. An InSb bolometer was used again as detector.
Further details on this spectrometer system are available in
\citet{COSSTA}.

The pressure of CO in the 3 and 2~m, respectively, long absorption cells
was in the $1 - 4$~Pa range. The measurements were carried out in a static
mode or in a slow flow through the cell. One has to keep in mind that
in the terahertz region a flow speed of several m/s causes frequency shifts
of a few tens of kilohertz.

The smaller dipole moment in $\varv = 1$ compared with the vibrational 
ground state reduces the intensities by about 40\,\%; in $\varv = 3$ 
the intensity is decreased by almost one order of magnitude, 
see Fig.~\ref{dip}. 
The high excitation energy reduces the intensities much more, 
by about 4.5 orders of magnitude for each vibrational quantum at
300~K. For these reasons, vibrationally excited CO could not be observed 
even under very favorable conditions and with long integration times 
at room temperature for $\varv > 1$.
A glow discharge effectively increased the vibrational temperature 
such that lines of vibrationally excited CO could be recorded 
below 1~THz with the CTS. The available source power at 2~THz is 
considerably lower with COSSTA. Therefore, a modulation of the 
discharge (on/off) at $\sim5$~Hz was employed in addition to the 
usual frequency modulation to reduced baseline effects and to 
improve the spectrometer stability.
The measurements were carried out similarly to those on HNC
\citep{HNC_rot_2000} and DNC \citep{DNC_rot_2006} in their ground
and $\varv _2 = 1$ excited vibrational states.

A high accuracy of a transition frequency in our measurements 
depends very much on the flatness of the baseline, the absence of other, 
interfering lines in the vicinity, and on a large signal-to-noise ratio 
(S/N). Relative accuracies of about $10^{-8}$ have been reached for 
strong, isolated lines recorded in Doppler mode up to 1~THz as well 
as near 2~THz fairly commonly as demonstrated in the investigations 
of, e.g., H$_2$CO \citep{H2CO_2003} and SO$_2$ in its $\varv _2 = 0$ 
and 1 vibrational states \citep{SO2_2005}. Under favorable conditions, 
better accuracies can be reached even in Doppler-limited measurements, 
as shown in the studies of HCN, $\varv _2 = 1$ \citep{HCN_lab_2003}, 
and SiS \citep{SiS_2007}. The study of rotational transitions of CO 
by \citet{CO_TuFIR} demonstrates that such high relative accuracies 
are possible also in other laboratories and at even higher frequencies 
(beyond 4~THz); see also \citet{CO_v0_rot_1997} for an updated list 
of these frequencies and uncertainties. 
It is worthwhile mentioning that in sub-Doppler measurements 
relative accuracies of better than $10^{-9}$ have been reached e.g. 
in measurements of CO \citep{CO_v0_rot_1997} 
and $^{13}$CO \citep{1316_GC}.

\section{Analysis and discussion}
\label{a&d}


\begin{figure}
\includegraphics[angle=-90,width=9cm]{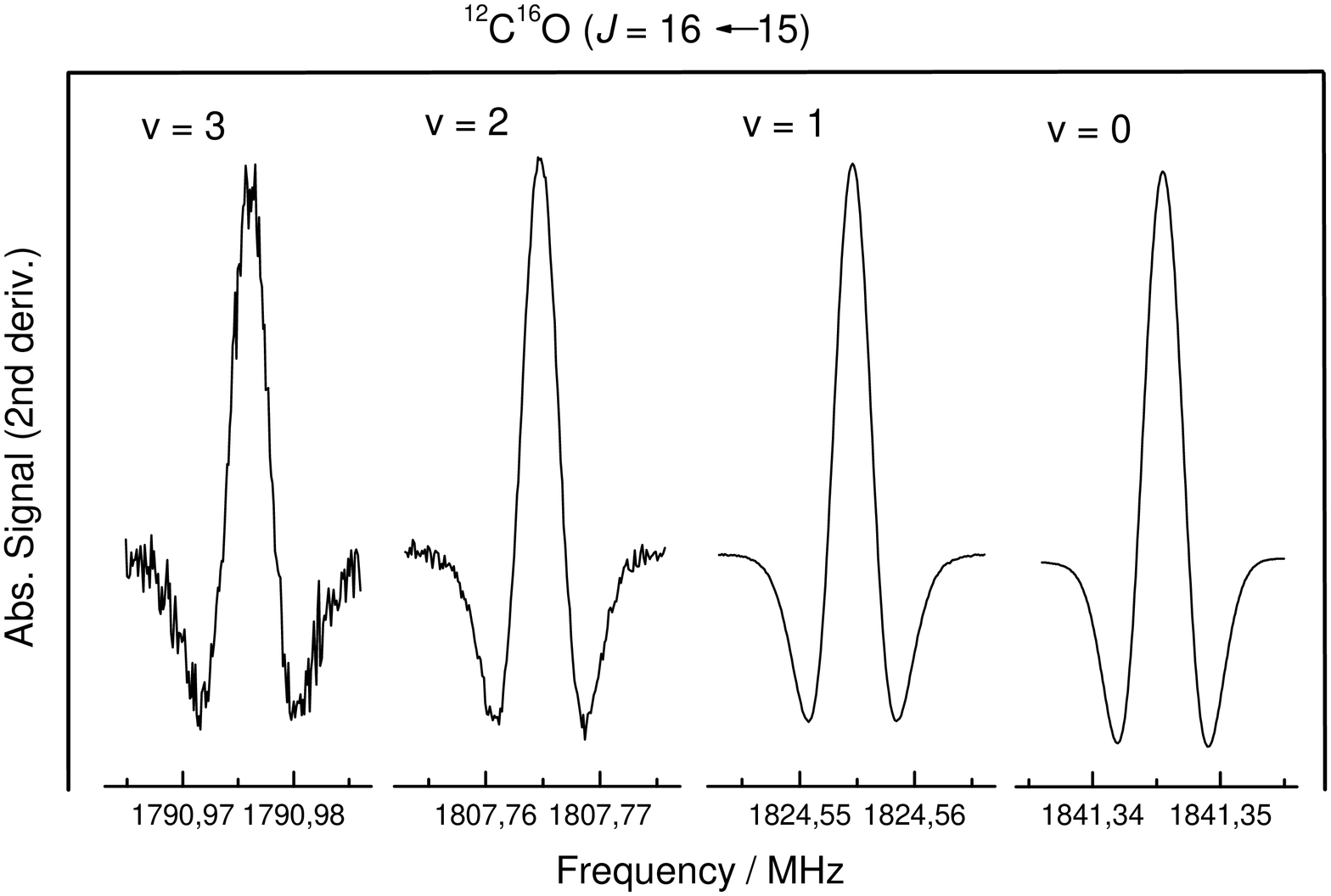}
  \caption{The $J = 16 \gets 15$ transitions are shown for $\varv = 0$ 
   to 3. Note that the signal-to-noise ratio is very good even in the 
   $\varv = 3$ state. The lower state energy of the transition is 
   6800~cm$^{-1}$, see Table~\ref{lines}. The lines appear as second 
   derivatives of a Gaussian line-shape because of the 2$f$-modulation.}
\label{spektrum}
\end{figure}


Predictions of the CO, $\varv = 1$ to 3 rotational spectrum were based
initially on infrared measurements as summarized, e.g. in
\citet{CO_pot_2004}.
All but the very weak $J = 1 - 0$ transition were recorded for $\varv = 1$ 
below 1~THz, fewer transitions were recorded for higher excited states.
Two additional transitions could be recorded for each vibrational state 
with COSSTA near 2~THz. Fig.~\ref{spektrum} displays 
the very good signal-to-noise ratios obtained for the $\varv = 0$ to 
3 states.  
Altogether between 9 and 4 experimental transition frequencies were 
obtained in the course of the present investigations. They 
are given in Table~\ref{lines} together with their uncertainties. 
Also included are the calculated frequencies up to $J = 20 - 19$ with 
uncertainties, the lower state energies, and the line strengths $S\mu ^2$.

The spectroscopic parameters have been determined in a weighted
least-squares fit, and are listed in Table~\ref{parameter} 
together with those of the ground vibrational state from 
\citet{CO_v0_rot_1997}. The parameters $B_{\varv}$ and
$D_{\varv}$ have been determined very well. Even the sextic
centrifugal distortion term $H_{\varv}$ has a standard deviation 
that is less than 10\,\% of its value for $\varv = 1$, 
but it is barely determined for $\varv = 2$ and 3. 
The rotational constant $B_{\varv}$ decreases with vibrational 
excitation, as is generally the case. 
The change in $D_{\varv}$ is very small, but still determined with
significance; the latter is not the case for $H_{\varv}$.
The rms error of the fits, also known as the weighted standard
deviation of the fit, are 0.561 for $\varv = 1$ and 2 suggesting 
the error estimates of the rest frequencies to be conservative. 
The rms error of 0.2 for $\varv = 3$ is not meaningful because 3 
parameters have been determined from 4 transition frequencies only.

Table~\ref{lines} provides calculated rest frequencies up to 2.3~THz.
The small values of the predicted uncertainties suggest that reliable
predictions of the rotational spectrum of CO $\varv = 1$ and probably 2 
can be made up to at least 3~THz. 
Predictions to even higher frequencies will be available in the 
catalog section of the Cologne Database for Molecular
spectroscopy\footnote{website: http://www.astro.uni-koeln.de/cdms/}
\citep{CDMS_1,CDMS_2}.

\section{Astronomical outlook}

As mentioned above, pure rotational emission from the 
$\varv =1, J = 2 - 1$ and $3 - 2$ lines has very recently been detected 
towards the hot, innermost circumstellar envelope of the high mass-loss 
carbon AGB star IRC+10216 \citep{Patel2009a}. 
Potentially, other vibrationally excited rotational lines from CO are 
detectable toward this and similar objects as well. 
The observations just reported, made with the Submillimeter Array, 
have an angular resolution of a few arc seconds and by no means resolve 
the emitting region, which is estimated to have more than 10 times 
smaller dimensions, $\sim 0\rlap{.}''2$ of $\sim 30$ AU 
for a distance of 150 pc \citep{CrosasMenten1997}. 
Resolving such small regions is outside the capabilities of existing 
(sub)millimeter interferometers.

However, the Atacama Large Millimeter Array (ALMA), currently under 
construction on the $\sim 5000$ meter high Llano de Chajnantor in Chile, 
with its fifty 12 m diameter antennas will have a \textit{much} 
higher angular resolution and a \textit{much} larger collecting area 
than all existing instruments. 
Consequently, ALMA will have a vastly larger brightness temperature 
sensitivity, $\Delta T_{\rm B} \propto 
[N(N-1)A_{\rm eff}]^{-1}(\nu\theta_{\rm B})^{-2}(\Delta \nu)^{-1/2}$. 
Here $N$ is the number of antennas (currently 50 planned and 64 as a goal), 
$A_{\rm eff}$ the effective area of a single antenna, $\nu$ the 
observing frequency, $\theta_{\rm B}$ the synthesized beam size and 
$\Delta \nu$ the frequency resolution.

Using the online ALMA Sensitivity Calculator 
Tool\footnote{http://www.eso.org/sci/facilities/alma/observing/tools/etc/} 
on ESO's ALMA WWW site we employ inputs equivalent to $\nu = 340$ GHz 
(comparable to the frequencies of vibrationally excited $J = 3 - 2$ CO line) 
and $\Delta \nu = 1$~MHz (corresponding to 0.88~km~s$^{-1}$ velocity resolution; 
i.e., appropriate for the $\sim 7$~km~s$^{-1}$ width of the lines 
observed toward IRC+10216), we calculate for 1.2~mm precipitable water vapor 
(mediocre weather conditions for the ALMA site) an rms flux density sensitivity 
of 3.2~mJy for 1~h of integration time. 
In an extended configuration, with a maximum baseline length,  
$B_{\rm max}$, of 10~km, a beam of FWHM $\theta_{\rm B} = c/(\nu B_{\rm max})$ 
radians, or 22 milli arcseconds could be synthesized with ALMA 
(at $\nu = 340$ GHz; $c$ is the speed of light). 
With this  $\theta_{\rm B}$ and the flux density sensitivity quoted above, 
we calculate that an rms brightness temperature sensitivity of 70~K could be 
achieved with ALMA at that resolution, which corresponds to a quarter of the 
diameter of IRC+10216's radio photosphere (Menten et al. in prep.). 

Given that the brightness temperature of the gas emitting vibrationally excited CO 
is in high likelihood hotter than 1000~K \citep{Patel2009a},  
spatially resolved imaging will allow studies of the distribution and dynamics of the 
hot material close to the stellar atmosphere of IRC+10216 and similar evolved stars. 
This will provide important insight into the origins of the mass loss process 
in such objects. 
These high resolution observations that will also easily image the radio photospheres 
of such near-by red giant stars \citep[see, e.g., ][]{ReidMenten2007} 
will be eminently feasible since their stellar continuum emission will provide 
ample flux for self calibration.



\begin{acknowledgements}
This work has been supported by the Deutsche Forschungsgemeinschaft
(DFG) via the collaborative research grant SFB~494. H.S.P.M. is grateful
for recent support by the Bundesministerium f\"ur Bildung und Forschung
(BMBF) administered through Deutsches Zentrum f\"ur Luft- und Raumfahrt
(DLR). His support was aimed in particular at maintaining the CDMS.
\end{acknowledgements}



\newpage
\begin{table*}
\caption{Measured transition frequencies$^a$ (MHz) of vibrationally
excited CO, transition frequencies$^a$ (MHz) calculated from the final
set of spectroscopic parameters, lower state energies $E_{\rm lo}$
(cm$^{-1}$) and line strengths $S\mu ^2$ (10$^{-3}$~D$^2$)}
\label{lines}
\renewcommand{\arraystretch}{0.65}
\begin{tabular}{rr@{}lr@{}lr@{}lr@{}l}
\hline
 & \multicolumn{8}{c}{$\varv = 1$} \\
\hline
$J''$ &\multicolumn{2}{c}{measured} &\multicolumn{2}{c}{calculated}
&\multicolumn{2}{c}{$E_{\rm lo}$} &\multicolumn{2}{c}{$S\mu ^2$} \\
\hline
 0 &          --&          &    114\,221&.7523~(13) & 2143&.271 &   7&.228 \\
 1 &    228\,439&.074~(25) &    228\,439&.1008~(24) & 2147&.081 &  14&.444 \\
 2 &    342\,647&.636~(20) &    342\,647&.6421~(33) & 2154&.701 &  21&.638 \\
 3 &    456\,842&.977~(10) &    456\,842&.9726~(37) & 2166&.131 &  28&.795 \\
 4 &    571\,020&.677~(12) &    571\,020&.6895~(37) & 2181&.369 &  35&.909 \\
 5 &    685\,176&.392~(5)  &    685\,176&.3902~(34) & 2200&.416 &  42&.960 \\
 6 &    799\,305&.677~(10) &    799\,305&.6731~(32) & 2223&.271 &  49&.950 \\
 7 &    913\,404&.136~(5)  &    913\,404&.1370~(40) & 2249&.933 &  56&.851 \\
 8 &          --&          & 1\,027\,467&.3817~(59) & 2280&.401 &  63&.657 \\
 9 &          --&          & 1\,141\,491&.0083~(84) & 2314&.674 &  70&.366 \\
10 &          --&          & 1\,255\,470&.619~(11)  & 2352&.750 &  76&.954 \\
11 &          --&          & 1\,369\,401&.816~(13)  & 2394&.628 &  83&.431 \\
12 &          --&          & 1\,483\,280&.205~(15)  & 2440&.306 &  89&.760 \\
13 &          --&          & 1\,597\,101&.393~(15)  & 2489&.783 &  95&.941 \\
14 &          --&          & 1\,710\,860&.986~(13)  & 2543&.057 & 101&.97  \\
15 & 1\,824\,554&.595~(10) & 1\,824\,554&.5953~(83) & 2600&.125 & 107&.85  \\
16 & 1\,938\,177&.832~(10) & 1\,938\,177&.8318~(92) & 2660&.986 & 113&.51  \\
17 &          --&          & 2\,051\,726&.309~(23)  & 2725&.636 & 119&.04  \\
18 &          --&          & 2\,165\,195&.644~(46)  & 2794&.074 & 124&.33  \\
19 &          --&          & 2\,278\,581&.453~(77)  & 2866&.298 & 129&.41  \\
\hline
 & \multicolumn{8}{c}{$\varv = 2$} \\
\hline
 0 &          --&          &    113\,172&.3761~(31) & 4260&.062 &  3&.588 \\
 1 &    226\,340&.341~(20) &    226\,340&.3489~(56) & 4263&.837 &  7&.169 \\
 2 &    339\,499&.521~(20) &    339\,499&.5151~(72) & 4271&.387 & 10&.732 \\
 3 &          --&          &    452\,645&.4717~(76) & 4282&.712 & 14&.270 \\
 4 &    565\,773&.818~(10) &    565\,773&.8161~(69) & 4297&.810 & 17&.778 \\
 5 &    678\,880&.140~(10) &    678\,880&.1462~(66) & 4316&.682 & 21&.240 \\
 6 &    791\,960&.061~(25) &    791\,960&.0603~(98) & 4339&.327 & 24&.658 \\
 7 &    905\,009&.183~(25) &    905\,009&.158~(16)  & 4365&.744 & 28&.017 \\
 8 &          --&          & 1\,018\,023&.038~(25)  & 4395&.932 & 31&.304 \\
 9 &          --&          & 1\,130\,997&.302~(34)  & 4429&.890 & 34&.525 \\
10 &          --&          & 1\,243\,927&.553~(42)  & 4467&.616 & 37&.665 \\
11 &          --&          & 1\,356\,809&.392~(48)  & 4509&.109 & 40&.719 \\
12 &          --&          & 1\,469\,638&.425~(50)  & 4554&.367 & 43&.677 \\
13 &          --&          & 1\,582\,410&.258~(45)  & 4603&.389 & 46&.526 \\
14 &          --&          & 1\,695\,120&.497~(32)  & 4656&.172 & 49&.273 \\
15 & 1\,807\,764&.748~(15) & 1\,807\,764&.752~(18)  & 4712&.716 & 51&.907 \\
16 & 1\,920\,338&.703~(80) & 1\,920\,338&.633~(48)  & 4773&.016 & 54&.408 \\
17 &          --&          & 2\,032\,837&.75~(11)   & 4837&.072 & 56&.797 \\
18 &          --&          & 2\,145\,257&.73~(21)   & 4904&.880 & 59&.032 \\
19 &          --&          & 2\,257\,594&.17~(33)   & 4976&.438 & 61&.146 \\
\hline
 & \multicolumn{8}{c}{$\varv = 3$} \\
\hline
 0 &          --&          &    112\,123&.087~(12) & 6350&.439 &   1&.207 \\
 1 &          --&          &    224\,241&.770~(22) & 6354&.179 &   2&.410 \\
 2 &          --&          &    336\,351&.644~(30) & 6361&.659 &   3&.603 \\
 3 &          --&          &    448\,448&.307~(35) & 6372&.879 &   4&.781 \\
 4 &          --&          &    560\,527&.354~(37) & 6387&.837 &   5&.939 \\
 5 &    672\,584&.390~(40) &    672\,584&.384~(37) & 6406&.534 &   7&.075 \\
 6 &          --&          &    784\,614&.995~(40) & 6428&.969 &   8&.182 \\
 7 &    896\,614&.750~(100)&    896\,614&.785~(49) & 6455&.141 &   9&.255 \\
 8 &          --&          & 1\,008\,579&.357~(65) & 6485&.049 &  10&.290 \\
 9 &          --&          & 1\,120\,504&.311~(85) & 6518&.692 &  11&.284 \\
10 &          --&          & 1\,232\,385&.25~(11)  & 6556&.068 &  12&.232 \\
11 &          --&          & 1\,344\,217&.78~(12)  & 6597&.176 &  13&.132 \\
12 &          --&          & 1\,455\,997&.51~(13)  & 6642&.014 &  13&.976 \\
13 &          --&          & 1\,567\,720&.05~(12)  & 6690&.581 &  14&.762 \\
14 &          --&          & 1\,679\,381&.012~(94) & 6742&.874 &  15&.488 \\
15 & 1\,790\,976&.008~(50) & 1\,790\,976&.004~(49) & 6798&.892 &  16&.151 \\
16 & 1\,902\,500&.636~(100)& 1\,902\,500&.647~(96) & 6858&.633 &  16&.747 \\
17 &          --&          & 2\,013\,950&.56~(25)  & 6922&.094 &  17&.275 \\
18 &          --&          & 2\,125\,321&.36~(47)  & 6989&.272 &  17&.728 \\
19 &          --&          & 2\,236\,608&.68~(78)  & 7060&.165 &  18&.110 \\
\hline
\end{tabular}\\[2pt]

$^a$ Numbers in parentheses are one standard deviation in units of the
least significant figures.
\end{table*}


\begin{table*}
\caption{Spectroscopic parameters$^a$ (MHz) of CO in excited
vibrational states compared with that in the ground vibrational state}
\label{parameter}
\renewcommand{\arraystretch}{1.10}
\begin{tabular}{rr@{}lr@{}lr@{}lr@{}l}
\hline
Parameter & \multicolumn{2}{c}{$\varv = 0^b$} & \multicolumn{2}{c}{$\varv = 1$} & \multicolumn{2}{c}{$\varv = 2$} & \multicolumn{2}{c}{$\varv = 3$} \\
\hline
$B_{\varv}$             & 57\,635&.968\,019~(28) & 57\,111&.243\,12~(66) & 56\,586&.555\,00~(157) & 56\,061&.910\,5~(59) \\
$D_{\varv} \times 10^3$ &     183&.504\,89~(16)  &     183&.490\,3~(71)  &     183&.474\,0~(234)  &     183&.511~(63)    \\
$H_{\varv} \times 10^9$ &     171&.68~(10)       &     177&.9~(145)      &     163&.~(53)         &     241&.~(134)      \\
\hline
\end{tabular}\\[2pt]
$^a$ Numbers in parentheses are one standard deviation in units of the
least significant figures.\\
$^b$ \citet{CO_v0_rot_1997}
\end{table*}


\begin{thebibliography}{}

\bibitem[Ag{\'u}ndez et al.(2007)]{CS_v_3}
Ag{\'u}ndez, M., Cernicharo, J., \& Gu{\'e}lin, M.
2007, \apjl, 662, L91

\bibitem[Ag{\'u}ndez et al.(2008)]{SiS_v_4}
Ag{\'u}ndez, M., Cernicharo, J., Pardo, J.~R., Gu{\'e}lin, M.,
\& Phillips, T.~G.
2008, \aap, 485, L33

\bibitem[Bensch et al.(2001)]{Det_13C17O}
Bensch, F., Pak, I., Wouterloot, J.~G.~A., Klapper, G.,
\& Winnewisser, G.
2001, \apjl, 562, L185

\bibitem[Br{\"u}nken et al.(2003)]{H2CO_2003} 
Br{\"u}nken, S., M{\"u}ller, H.~S.~P., Lewen, F., \& Winnewisser, G. 
2003, Phys. Chem. Chem. Phys., 5, 1515 

\bibitem[Br{\"u}nken et al.(2006)]{DNC_rot_2006}
Br{\"u}nken, S., M{\"u}ller, H.~S.~P., Thorwirth, S., Lewen, F.,
\& Winnewisser, G.
2006, J. Mol. Struct., 780, 3

\bibitem[Bogey et al.(1986)]{CO_vib_J_2-1}
Bogey, M., Demuynck, C., Destombes, J.~L., \& Lapauw, J.~M.
1986, J. Phys. E, 19, 520

\bibitem[Buhl et al.(1974)]{SiO_v_2}
Buhl, D., Snyder, L.~E., Lovas, F.~J., \& Johnson, D.~R.
1974, \apjl, 192, L97

\bibitem[Cazzoli et al.(2002)]{1217_GC}
Cazzoli, G., Dore, L., Puzzarini, C., \& Beninati, S.
2002, Phys. Chem. Chem. Phys., 4, 3575

\bibitem[Cazzoli et al.(2003)]{1218_GC}
Cazzoli, G., Puzzarini, C., \& Lapinov, A.~V.
2003, \apjl, 592, L95

\bibitem[Cazzoli et al.(2004)]{1316_GC}
Cazzoli, G., Puzzarini, C., \& Lapinov, A.~V.
2004, \apj, 611, 615

\bibitem[Cernicharo et al.(1993)]{SiO_v_4}
Cernicharo, J., Bujarrabal, V., \& Santaren, J.~L.
1993, \apjl, 407, L33

\bibitem[Cherchneff \& Barker(1992)]{Cherchneff1992} 
Cherchneff, I., \& Barker, J.~R. 
1992, \apj, 394, 703

\bibitem[Coxon \& Hajigeorgiou(2004)]{CO_pot_2004}
Coxon, J.~A., \& Hajigeorgiou, P.~G.
2004, \jcp, 121, 2992

\bibitem[Crosas \& Menten(1997)]{CrosasMenten1997} 
Crosas, M., \& Menten, K.~M. 
1997, \apj, 483, 913

\bibitem[Cummins et al.(1980)]{Cummins1980} 
Cummins, S.~E., Morris, M., \& Thaddeus, P. 
1980, \apj, 235, 886

\bibitem[Gendriesch et al.(2000)]{COSSTA}
Gendriesch, R., Lewen, F., Winnewisser, G., \& Hahn, J.
2000, J. Mol. Spectrosc., 203, 205

\bibitem[Gonzalez-Alfonso et al.(1996)]{29SiO_v_3}
Gonzalez-Alfonso, E., Alcolea, J., \& Cernicharo, J.
1996, \aap, 313, L13

\bibitem[Goorvitch(1994)]{CO_dipole_surface}
Goorvitch, D.
1994, \apjs, 95, 535

\bibitem[Keady et al.(1988)]{Keady1988} 
Keady, J.~J., Hall, D.~N.~B., \& Ridgway, S.~T. 
1988, \apj, 326, 832

\bibitem[Klapper et al.(2000a)]{1316_GK}
Klapper, G., Lewen, F., Gendriesch, R., Belov, S.~P.,
\& Winnewisser, G.
2000, J. Mol. Spectrosc., 201, 124

\bibitem[Klapper et al.(2000b)]{1318_GK}
Klapper, G., Lewen, F., Belov, S.~P., \& Winnewisser, G.
2000, Z. Naturforsch., 55a, 441

\bibitem[Klapper et al.(2001)]{1218_GK}
Klapper, G., Lewen, F., Gendriesch, R., Belov, S.~P.,
\& Winnewisser, G.
2001, Z. Naturforsch., 56a, 329

\bibitem[Klapper et al.(2003)]{1217_1317_GK}
Klapper, G., Surin, L., Lewen, F., M{\"u}ller, H.~S.~P.,
Pak, I., \& Winnewisser, G.
2003, \apj, 582, 262

\bibitem[Lafont et al.(1982)]{Lafont1982} 
Lafont, S., Lucas, R., \& Omont, A. 
1982, \aap, 106, 201

\bibitem[M{\"u}ller et al.(2001)]{CDMS_1}
M{\"u}ller, H.~S.~P., Thorwirth, S., Roth, D.~A.,
\& Winnewisser, G.
2001, A\&A, 370, L49-L52

\bibitem[M{\"u}ller et al.(2005)]{CDMS_2}
M{\"u}ller, H.~S.~P., Schl{\"o}der, F., Stutzki, J.,
\& Winnewisser, G.
2005, J. Mol. Struct, 742, 215

\bibitem[M{\"u}ller \& Br{\"u}nken(2005)]{SO2_2005} 
M{\"u}ller, H.~S.~P., \& Br{\"u}nken, S. 
2005, J. Mol. Spectrosc., 232, 213 

\bibitem[M{\"u}ller et al.(2007)]{SiS_2007}
M{\"u}ller, H.~S.~P., et al.
2007, Phys. Chem. Chem. Phys., 9, 1579

\bibitem[Muenter(1975)]{CO_Stark}
Muenter, J.~S.
1975, J. Mol. Spectrosc., 55, 490

\bibitem[Patel et al.(2009a)]{Patel2009a} 
Patel, N.~A., Young, K.~H., Br{\"u}nken, S., Menten, 
K.~M., Thaddeus, P., \& Wilson, R.~W. 
2009, \apjl, 691, L55 

\bibitem[Patel et al.(2009b)]{Patel2009b} 
Patel, N.~A., et al.
2009b, \apj, accepted, arXiv:0811.2142

\bibitem[Puzzarini et al.(2003)]{1317_1318_CP}
Puzzarini, C., Dore, L., \& Cazzoli, G.,
2003, J. Mol. Spectrosc., 217, 19

\bibitem[Reid \& Menten(2007)]{ReidMenten2007} 
Reid, M.~J., \& Menten, K.~M.\ 
2007, \apj, 671, 2068

\bibitem[Sanz et al.(2003)]{SiO_2003}
Sanz, M.~E., McCarthy, M.~C., \& Thaddeus, P.
2003, \jcp, 119, 11715

\bibitem[Sahai \& Wannier(1985)]{SahaiWannier1985} 
Sahai, R., \& Wannier, P.~G. 
1985, \apj, 299, 424

\bibitem[Scoville \& Solomon(1978)]{ScovilleSolomon1978}
Scoville, N.~Z., \& Solomon, P.~M. 
1978, \apjl, 220, L103

\bibitem[Snyder \& Buhl(1974)]{SiO_vib_1}
Snyder, L.~E., \& Buhl, D.
1974, \apjl, 189, L31

\bibitem[Thorwirth et al.(2000)]{HNC_rot_2000}
Thorwirth, S., M{\"u}ller, H.~S.~P., Lewen, F., Gendriesch, R.,
\& Winnewisser, G.
2000, \aap, 363, L37

\bibitem[Thorwirth et al.(2003)]{HCN_lab_2003} 
Thorwirth, S., M{\"u}ller, H.~S.~P., Lewen, F., Br{\"u}nken, 
S., Ahrens, V., \& Winnewisser, G. 
2003, \apjl, 585, L163 

\bibitem[Tsuji(1964)]{Tsuji1964} 
Tsuji, T. 
1964, Ann. Tokyo Astron. Obs., 9, 00

\bibitem[Turner(1987a)]{CS_vib}
Turner, B.~E.
1987a, \aap, 182, L15

\bibitem[Turner(1987b)]{SiS_vib}
Turner, B.~E.
1987b, \aap, 183, L23

\bibitem[Varberg \& Evenson(1992)]{CO_TuFIR} 
Varberg, T.~D., \& Evenson, K.~M. 
1992, \apj, 385, 763 

\bibitem[Winnewisser et al.(1994)]{BWO-THz_spec}
Winnewisser, G., et al.
1994, J. Mol. Spectrosc., 165, 294

\bibitem[Winnewisser et al.(1997)]{CO_v0_rot_1997}
Winnewisser, G., Belov, S.~P., Klaus, T., \& Schieder, R.
1997, J. Mol. Spectrosc., 184, 468

\bibitem[Zink et al.(1990)]{1316_TuFIR}
Zink, L.~R., de Natale, P., Pavone, F.~S., Prevedelli, M.,
Evenson, K.~M., \& Inguscio, M.
1990, J. Mol. Spectrosc., 143, 304

\bibitem[Zapata et al.(2009)]{Zapata2009} 
Zapata, L.~A., Menten, K., Reid, M., \& Beuther, H. 
2009, \apj, 691, 332 

\end{thebibliography}
\end{document}